\documentclass[useAMS,usenatbib,usegraphicx]{mn2e}

\newcommand{\Kep}{{\it Kepler~}}
\newcommand{\cd}{d$^{-1}$}
\newcommand{\kms}{km\,s$^{-1}$}
\newcommand{\vsini}{$v$\,sin\,$i$}

\title[Discovery of a magnetic field in the $\delta$\,Scuti star
HD\,188774]{First discovery of a magnetic field in a main sequence 
$\delta$\,Scuti star: the \Kep star HD\,188774\thanks{Based on
observations obtained at the Canada-France-Hawaii Telescope (CFHT) operated by
the National Research Council of Canada, the Institut National des Sciences de
l'Univers of the CNRS of France, and the University of Hawaii.}}
\author[C. Neiner and P. Lampens]{C. Neiner$^{1}$\thanks{E-mail:
coralie.neiner@obspm.fr} and P. Lampens$^{2}$\\
$^{1}$LESIA, Observatoire de Paris, PSL Research University, CNRS,
Sorbonne Universit\'es, UPMC Univ. Paris 06, Univ. Paris Diderot,\\
Sorbonne Paris Cit\'e, 5 place Jules Janssen, 92195 Meudon, France\\
$^{2}$Koninklijke Sterrenwacht van Belgi\"e, Ringlaan 3, 1180, Brussel, Belgium}
\begin{document}

\date{Accepted ... Received ...}

\pagerange{\pageref{firstpage}--\pageref{lastpage}} \pubyear{2015}

\maketitle

\label{firstpage}

\begin{abstract}
The \Kep space mission provided a wealth of $\delta$\,Sct-$\gamma$\,Dor  hybrid 
candidates. While some may be genuine hybrids, others might be misclassified due
to the presence of a binary companion or to rotational modulation caused by
magnetism and related surface inhomogeneities. In particular, the \Kep
$\delta$\,Sct-$\gamma$\,Dor hybrid candidate HD\,188774 shows a few low
frequencies in its light and radial velocity curves, whose origin is unclear. In
this work, we check for the presence of a magnetic field in HD\,188774. We
obtained two spectropolarimetric measurements  with ESPaDOnS at CFHT. The data
were analysed with the least squares deconvolution method. We detected a clear
magnetic signature in the Stokes $V$ LSD profiles. The origin of the low
frequencies detected in HD\,188774 is therefore most probably the rotational
modulation of surface spots possibly related to the presence of a magnetic
field. Consequently, HD\,188774 is not a genuine hybrid
$\delta$\,Sct-$\gamma$\,Dor star, but the first known magnetic main sequence
$\delta$\,Sct star. This makes it a prime target for future asteroseismic and
spot modelling. This result casts new light on the interpretation of the \Kep
results for other $\delta$\,Sct-$\gamma$\,Dor hybrid candidates.
\end{abstract}


\section{Introduction}\label{intro}

The region of the Hertzsprung-Russell (H-R) diagram where A and F  stars are
located shows a rich variety of stellar atmospheric processes, several of which
can produce short- and/or long-term variability. These processes consist in
different pulsation mechanisms, as well as various other phenomena involving
convection, diffusion, rotation, and even magnetic fields. Multiplicity, which
has a non-negligible impact on some  processes, is an additional cause for
variability. As an illustration, the binary fraction of a sample of A-F stars
from the Sco-Cen association is found to lie between 60 and 80\%
\citep{janson2013}.

The A and F stars define some of the most populated groups of pulsators: namely
the $\delta$ Scuti ($\delta$\,Sct) and $\gamma$\,Doradus ($\gamma$\,Dor) stars.
They have masses between 2.5 and 1.5 M$_{\odot}$. The $\delta$\,Sct and
$\gamma$\,Dor stars are found in neighbouring, even overlapping, regions of the
H-R diagram. The driving mechanism of the $\delta$\,Sct stars is the $\kappa$
mechanism operating in the He\,{\sc ii} ionisation region
\citep{pamyatnykh2000}, producing p modes. The driving mechanism of the $\gamma$
Dor pulsators is convective blocking near the base of their convective envelopes
\citep{guzik2000, dupret2004}, producing g modes. This mechanism can only
operate if the outer convective layer has a depth between  3 and 9\% of the
stellar radius. The overlapping region, where both types of pulsation modes can
co-exist, yields the potential to constrain physical properties in the various
internal layers of the star, from the core to the outer envelope. The hybrid
pulsators are thus very promising targets for asteroseismic studies.

Before the advent of asteroseismic space missions, only a few stars were known
to show characteristics of both pulsation types. The first hybrid $\gamma$
Dor-$\delta$  Sct star was discovered from the ground \citep{henry2005}.
Meanwhile, hundreds of candidate hybrid pulsators have been identified by
asteroseismic space missions. The first results based on data collected by the
\Kep mission \citep{grigahcene2010, uytterhoeven2011} suggest that the number
of  hybrid candidates is much higher than expected. However, the fact that many
are found across the full width of both observational instability strips poses a
serious  problem.  If the majority of the candidates were proven to be true
hybrids, current theory  would need to be revised \citep[e.g. confrontation of
Figs. 2 and 3 in][]{grigahcene2010}. The origin of the low frequencies ($<$ 5
c~d$^{-1}$) found in the periodograms of practically all of the \Kep
$\delta$\,Sct stars \citep{balona2014,balona2015}, often attributed to hybrid
pulsation, is thus puzzling, in particular for the hottest ones. Therefore, it
is very important to assess whether the hybrid candidates are all genuine hybrid
pulsators. 

Presently, about a dozen of genuine hybrids have been confirmed by detailed
investigations \citep[e.g.][]{hareter2012, tkachenko2013}, while the
classification of many new candidates rests on a ``first-look" photometric
analysis. A careful study of 69 candidate $\gamma$\,Dor stars with
spectroscopically determined atmospheric  parameters, of which 14 were
identified as $\gamma$ Dor-$\delta$ Sct hybrids, rather suggests that genuine
hybrids are confined to the effective temperature interval  [6900:7400] K, as
theory predicts \citep[see Fig.~15 in][]{dupret2005}.

Two other plausible scenarios can be considered to explain the presence of low
frequencies in $\delta$ Sct stars, without invoking $\gamma$ Dor pulsations:

First, pulsations can be superposed to the light curve of a binary star. In this
case, the object is an eclipsing or an ellipsoidal system (low frequencies) with
a $\delta$\,Sct component (high frequencies). Additionally, if the binary forms
a close system, pulsations may be affected by distorted stellar shapes and tidal
interactions \citep{reyniers2003b,reyniers2003a}. If the eccentricity becomes
significant, tidally excited g-modes may arise \citep[e.g.][]{hambleton2013}.
Another possibility consists in a binary with two components pulsating in
different frequency modes (i.e. a $\delta$\,Sct star with a $\gamma$\,Dor
companion).

Second, the object can be a $\delta$\,Sct star with some surface inhomogeneity.
In this case, the presence of large temperature gradients or abundance
variations (spots) on the stellar surface in combination with stellar rotation
induces low frequencies of the order of the rotational period (and its
harmonics) in the light curves  \citep{balona2011,balona2013}. 

In this Letter, we discuss and evaluate the various scenarios for the \Kep
$\delta$\,Sct-$\gamma$\,Dor candidate HD\,188774 (Sect.~\ref{star}). In
particular, we address the scenario of rotational modulation
(Sect.~\ref{rotmod}). We then present the spectropolarimetric results
(Sect.~\ref{obs}) and draw conclusions (Sect.~\ref{conclu}).

\section{HD\,188774}\label{star}

The \Kep $\delta$\,Sct-$\gamma$\,Dor (A7.5IV-III) hybrid candidate HD\,188774
(KIC\,5988140) was selected for a thorough study by \cite{lampens2013} 
\citep[see also][]{lampens2013_japan}, who also acquired 40 high-resolution
spectra with ground-based instruments. Fourier analysis of these data revealed
nine significant frequencies of the order of several hours, thus confirming the
$\delta$\,Sct oscillations. Furthermore, Fourier analysis of both light and
radial velocity (RV) curves revealed two additional low frequencies of higher
amplitude (f$_1$=0.68799 and f$_2$=0.343984 \cd).

At first sight, the \Kep light curve of HD\,188774 mimics an eclipsing binary
system with superposed short-period variations of $\delta$\,Sct type. Although
there is a strict 1:2 ratio between the dominant low frequencies corresponding
to the possible effects of ellipsoidality and reflection in the light curve,
\cite{lampens2013} ruled out the explanation of orbital motion because the RV
curve almost perfectly matches the variability pattern of the mean light curve
(after prewhitening of all the high frequencies).

In addition, with respect to the scenario of genuine hybrid pulsation, we find
several counter-arguments. Firstly, the exact integer ratio 1:2 between the two
most dominant low frequencies, the detection of additional harmonics of the
lowest frequency, and the phase relation between the two curves are atypical for
non-radial g-mode pulsations. Moreover, \cite{lampens2013} noted that the
light-to-velocity amplitude ratio is also very unusual: the observed amplitude
ratio is of the order of 0.63 mmag/\kms, whereas e.g. \cite{Aerts2004} derived a
mean ratio of about 15 mmag/\kms\ from an asteroseismic study of two bright
$\gamma$\,Dor stars. In addition, the effective temperature of HD\,188774
\citep[7600 $\pm$ 30 K,][]{lampens2013} is outside the temperature interval for
genuine hybrid stars.

Since both binarity and g-mode pulsations seem very unlikely, we investigated
below the third scenario, i.e. rotational modulation, as an explanation for the
low frequencies of variations observed in HD\,188774.

\section{Rotational modulation in HD\,188774}\label{rotmod}

Although A-type stars usually have homogeneous stellar surfaces due to their
large radiative envelopes with only shallow subsurface convection, a few do
present observable inhomogeneous surface distributions.  There are two possible
reasons:

(1) This can be due to (very) rapid rotation inducing large temperature
gradients \citep[e.g. Altair;][]{peterson2006}. In the case of HD\,188774, one
finds that critical rotation is highly improbable. Indeed, adopting the radius
listed in the \Kep Input Catalog (KIC) and assuming that P$_2$ = 2.90711~d from
\cite{lampens2013} is the rotation period, one finds that  $V_{\rm eq} =
62.3$~\kms. A comparison with the measured \vsini\, gives a probable inclination
angle of $i\sim50$\degr. In addition, \cite{lampens2013} found no temperature
variations between spectra taken at four different rotational phases.

\begin{table*}
\caption{Spectropolarimetric measurements of HD\,188774. The dates,
heliocentric Julian dates corresponding to the middle epoch of the
measurements, and exposure times are given. The computed longitudinal
field $B_l$ and $N$ values, with their respective error bars $\sigma$ and
significance level $z$ are also shown, as well as the field detection
probability in \% and in terms of type of detection.}
\label{tableBl}
\begin{tabular}{lllllllll}
\hline
Date & Mid-HJD & $T_{\rm exp}$ & $B_l \pm \sigma_B$ & $z_B$ & $N \pm \sigma_N$ & $z_N$ & Prob. & Detect. \\
 & -2450000 & s & G & & G & & \% & \\
\hline
Sep 7, 2014	& 2456907.951 & 4$\times$840 & 23.2 $\pm$ 17.1& 1.4 & 6.9 $\pm$ 17.1 & 0.4 & 99.999\% & Definite \\
Jul 23, 2015	& 2457227.027 & 10$\times$4$\times$129& 75.8 $\pm$ 13.0 & 5.8 & 7.6 $\pm$ 12.9 & 0.6 & 100\% & Definite \\
\hline
\end{tabular}
\end{table*}

(2) Another possible cause is local chemical inhomogeneities called ``spots"
\citep[e.g.][]{lehmann2006,lueftinger2010,kochukhov2011}. For HD\,188774,
\cite{lampens2013} used two simple models with symmetrically located spots to
represent as closely as possible the behaviour of the RV curve (cf. their
Fig.~10). Both models predict a peak-to-peak light amplitude of the order of
20\%, while the total amplitude of the \Kep light curve is only 0.5\% (i.e. 5
ppt). Because of this qualitative disagreement, they discarded the model of a
spotted surface for HD\,188774 as the explanation for the observed variations.
However, more  sophisticated spot models, which might explain the observed
variations, were not tested.

A major issue with the scenario of chemical spots is to be able to identify a
possible physical cause to explain their presence. Single A stars, in general,
do not have chemically spotted surfaces, unless they possess a magnetic field.
This is the case for most CP stars. In particular, Ap stars host very strong
magnetic fields \citep[e.g.][]{mathys2001}, which  break stellar rotation and
stabilize the atmosphere, enabling processes of atomic diffusion
\citep[e.g.][]{alecian2013}. Most of the He-strong and He-weak stars are also
known to host magnetic fields, and their He peculiarity is interpreted both in
terms of elemental diffusion and of the fractionation of their stellar
magnetized wind \citep{hunger1999}. In addition, recent results show that Am
stars can possess ultra-weak fields \citep{petit2011, blazere2014}. While their
low rotational velocity is usually attributed to their binary nature, their
magnetic fields may also contribute to enabling the diffusion process. On the
other hand, no HgMn stars have been found to host a magnetic field so far, in
spite of their chemical peculiarity  \citep[e.g.][]{makaganiuk2011}. Finally,
there seems to be no known A star with a fossil magnetic field of average
strength: either the fields are strong (B$_l >$100 G), e.g. in Ap stars, or they
are ultra weak (B$_l <$ 1 G), e.g. in Am stars. This phenomenon is referred to
as the magnetic dichotomy in intermediate-mass stars
\citep{auriere2007,lignieres2014}.

The abundance analysis of HD\,188774 did not show any chemical peculiarity, nor
abundance variations across the stellar surface \citep{lampens2013}. It is
nevertheless possible that this $\delta$\,Sct star hosts a fossil magnetic
field, as seen in 10\% of hot stars \citep{grunhut2015}. The field could be too
weak to produce chemical peculiarities or the pulsations could dominate the
field in terms of internal mixing. Moreover, \cite{lampens2013} demonstrated an
extremely stable \Kep light curve over a period of (at least) T = 682 days,
indicating that, if spots are present on the surface of HD\,188774, they should
remain stable over a few years. A fossil magnetic field could explain such
long-term spot stability.

A direct detection of a magnetic field has already been obtained in at least
one  non-peculiar A star: Vega \citep{lignieres2009, petit2010}. This star is a
very rapid rotator seen pole-on and hosts an ultra-weak field (B$_l$ = 0.6 G)
located in a spot near the rotation pole. In addition,
\cite{boehm2012,boehm2015} reported very small amplitude stellar radial velocity
variations, associated to either pulsations or rotational modulation. Thus, they
showed that non-peculiar A-type stars can have structures observable at the
surface.

In addition, other indirect evidence for magnetism has been detected in some A
stars, in particular in X-rays \citep{schroeder2007}. This activity may be
related to a dynamo field, similar to cooler stars. However, no direct detection
of a magnetic field has ever been obtained in these  active A stars.

Finally, magnetic A stars can undergo pulsations. In particular, very fast
oscillations occur in some Ap stars \citep[the roAp stars,
see][]{mathys1997,kochukhov2008}. However, no $\delta$\,Sct pulsations  have
been detected in a magnetic star so far, in spite of some attempts:
\cite{kurtz2008} claimed the first observational evidence for $\delta$\,Sct
pulsations and magnetic field in an Ap star \citep[HD\,21190; see
also][]{balona2011_roap}. However, \cite{bagnulo2012} showed that this magnetic
detection was spurious and that the star is probably not an Ap star. Conversely,
\cite{alecianwade2013} claimed a possible magnetic detection in the
$\delta$\,Sct star HD\,35929. However, their 5 spectropolarimetric measurements
gave contrasting results and this star is a pre-main sequence Herbig Ae star.

\section{Spectropolarimetric observations}\label{obs}

We obtained two circular spectropolarimetric measurements of HD\,188774 with
ESPaDOnS at CFHT. The first one was obtained on September 7, 2014 to test the
hypothesis of the presence of a magnetic field in this star. This measurement
consists of 4 subexposures obtained in different configurations of the
polarimeter. The exposure time was 4$\times$840 s, i.e. 3360 s in total. The
signal-to-noise  ratio (S/N) in the intensity spectrum $I$ peaks at 587. Using
the published RV ephemeris, our measurement corresponds to phase 0.189, close to
the minimum value of the RV curve \citep[see Fig.~3 in][]{lampens2013}. To
confirm the detection of a Zeeman signature in this first spectropolarimetric
measurement, a second measurement was acquired on July 23, 2015, which
corresponds to phase 0.950, close to the maximum value of the RV curve. This
measurement was obtained with 10 successive sequences of 4$\times$129 s each,
i.e. 5160 s in total, to make sure that the Zeeman signature is not polluted by 
the short-scale pulsations. See Table~\ref{tableBl}.

The spectra were reduced with the Libre-Esprit \citep{donati1997} and Upena
pipelines at CFHT. We obtained the intensity $I$ spectra, Stokes $V$ spectra, as
well as null $N$ polarisation spectra to check for spurious signatures. We
normalised the spectra to the continuum level with {\sc IRAF}.

Next, we applied the least squares deconvolution (LSD) method
\citep{donati1997}. This state-of-the-art technique is the most suited to
extract weak magnetic signatures in stellar spectra, as it allows to increase
the S/N of the magnetic signature by combining all available lines and weakens
at the same time signatures that would be present only in a few lines (thus
probably not of magnetic origin). The lines are combined using weights 
proportional to the line strength, wavelength $\lambda$, and Land\'e factor $g$.

The line mask used in these LSD calculations was extracted from the VALD
database \citep{piskunov1995, kupka1999}. We adopted the atmospheric parameters
T$_{\rm eff}$=7600 K and $\log g$=3.39 \citep{lampens2013}, assumed solar
abundances, and extracted only lines with a depth of 1\% or more of the
continuum level for the initial mask. From this line mask, we then rejected all
hydrogen lines, as well as lines blended with the H lines and/or contaminated by
telluric lines. The final mask contains 5260 lines, with an average Land\'e
factor of 1.191-1.192 and an average wavelength of 522.88-520.36 nm for the two
measurements, respectively. We then automatically adjusted the line depths to
provide the best fit to the observed Stokes $I$ spectra.

\begin{figure}
\begin{center}
\resizebox{0.8\hsize}{!}{\includegraphics[clip]{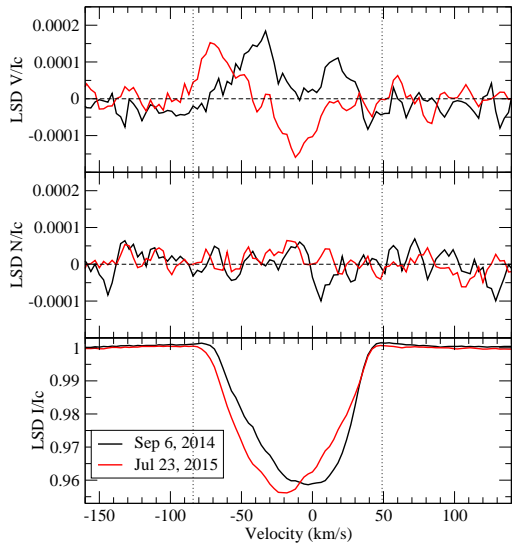}}
\caption[]{Stokes $V$ (top), $N$ (middle), Stokes $I$ (bottom) LSD profiles of
HD\,188774 for the first (black) and second (red) measurements. Vertical dotted
lines indicate the width of the profile over which the FAP and longitudinal
field were calculated.}
\label{HD188774LSD}
\end{center}
\end{figure}

For the second measurement, we co-added the 10 LSD profiles. The $I$, $V$ and
$N$ LSD profiles of both measurements are shown in Fig.~\ref{HD188774LSD}. The
S/N of the Stokes $I$ and $V$ LSD profiles is 1950 and 15240 for the first
measurement and 7520 and 19300 for the second measurement, respectively.

The $N$ profile is a combination of the spectropolarimetric subexposures in such
a way that the Zeeman signature of the stellar magnetic field cancels out
\citep{donati1997,bagnulo2009}. Only non-magnetic effects remain in the $N$
profile. The flat $N$ LSD profiles (Fig.~\ref{HD188774LSD}) show that the
measurements have not been contaminated by external sources, such as an
instrumental problem, although the first profile presents a small deviation from
the mean around 0 \kms. Since this measurement was obtained with longer
subexposures, corresponding to 1 to 23\% of the pulsation periods, it is
possible that pulsations have slightly influenced the profiles. In contrast, the
Stokes $V$ LSD profiles show clear Zeeman signatures.

We computed the False Alarm Probability (FAP) of a detection in the LSD Stokes
$V$ profiles inside the velocity range of the line ([-84:49] km~s$^{-1}$),
compared to the mean noise level outside the line. We used the convention
defined by \cite{donati1997}: if FAP $<$ 0.001\%, the magnetic detection is
definite, if 0.001\% $<$ FAP $<$ 0.1\% the detection is marginal, otherwise
there is no magnetic detection. Both measurements represent definite detections
of a magnetic field. On the contrary, no detection is found, neither outside the
line, nor in the $N$ profiles. Therefore, we can discard effects due to
instrumental origins or noise.

The two measurements show different shapes of the magnetic signatures. As
expected for a dipolar field, the first measurement obtained close to minimum RV
phase shows a symmetrical crossover signature, while the second measurement
taken close to maximum RV shows an asymmetrical signature (i.e. we see the
magnetic pole).

The shape of the magnetic signature in the first measurement may be reminiscent
of signatures observed in Am stars \citep[e.g.][]{blazere2014}. In these stars,
the positive double-lobe shape of the magnetic signatures is tentatively
explained by shocks in the thin convective layer \citep{blazere2015}. However,
it is more likely that this signature was obtained during a magnetic crossover
phase and that the central part of the signature does not reach negative values
because the profile is slightly modified by pulsations due to the long
subexposure time (as suggested by the $N$ profile). 

Next, we computed the longitudinal field strength $B_l$ from the Stokes $I$ and
$V$ LSD profiles \citep[following][]{rees1979,wade2000}, using the velocity
range between -84 and 49 km~s$^{-1}$. A similar calculation was performed using
the $N$ profile instead of $V$. The values of $B_l$, $N$, their error bars,  and
significance level are indicated in Table~\ref{tableBl}. As expected the first
crossover measurement corresponds to a $B_l$ close to 0, while the second
measurement is higher ($\sim$76 G). The higher significance level in $B_l$ than
in $N$ (see Table~\ref{tableBl}) confirms that the magnetic signature has a
stellar origin.

\section{Conclusions}\label{conclu}

The ESPaDOnS spectropolarimetric measurements of the \Kep hybrid candidate 
HD\,188774 exhibit definite magnetic signatures. Our result is the very first
detection of a magnetic field in a main sequence $\delta$\,Sct  pulsating star
and refutes the hybrid character of HD\,188774.

The possible existence of a dynamo magnetic field is unlikely, since most normal
A-type stars are found to be inactive: they are generally X-ray poor
\citep[$\sim$85\% of the sample observed by][]{schroeder2007} and only very
exceptionally do they also show flares in their light curves \citep{balona2012}.
On the contrary, fossil magnetic fields are found in $\sim$10\% of hot stars
\citep{grunhut2015}. Most of these magnetic A stars are the well-known Ap stars,
but a small fraction of them could be normal A stars, and a small fraction of
those could host $\delta$\,Sct pulsations. HD\,188774 is the first such example
to be discovered.

The observed relatively simple Zeeman signatures indeed  point to a possible
fossil origin of the field (as for OB and Ap stars), rather than to a dynamo.
This is also consistent with the long-term stability of the proposed spots at
the surface of HD\,188774, verified in the \Kep light curve. In addition, our
measurements of the longitudinal magnetic field of HD\,188774 are below 100 G,
which appears to lie on the edge of the dichotomy desert between strong and
ultra-weak fossil fields, as defined by \cite{auriere2007}. However, such values
suggest a polar field strength of a few hundred gauss, which is typical of
fossil fields in non-peculiar OB stars. 

Although \cite{lampens2013} considered the presence of spots as unlikely,
especially because of the small light-to-velocity amplitude ratio, the detection
of a magnetic field in HD\,188774 now makes this explanation to the observed low
frequencies of variations most plausible. Further spectropolarimetric
observations of this star covering a full rotation period should allow us to
constrain the magnetic field configuration and polar field strength.

In addition, the analysis of the pulsational properties of HD\,188774 based on
the \Kep light curve revealed nine significant $\delta$\,Sct-type frequencies
\citep{lampens2013}. Since we now know that the star possesses a magnetic field,
HD\,188774 is a very promising target for a detailed seismic and spotted surface
modelling, as well as for a study of the impact of the magnetic field on its
internal structure, mixing, convection, and oscillations.

HD\,188774 might well be the first member of a new class of weakly magnetic
$\delta$\,Sct stars. (Some of) the low frequencies observed in the light curves
of more candidate hybrid $\delta$\,Sct-$\gamma$\,Dor stars might be due to
rotational modulation induced by spots on their stellar surface, i.e. that they
might also turn out to be magnetic $\delta$\,Sct stars rather than hybrid stars.
This casts new light on the analysis and interpretation of the \Kep results for
the diversified and intriguing class of A-type stars.

\section*{Acknowledgments}

We thank the CFHT director, Doug Simons, for allocating the time to perform the
first observation presented here, and Jason Grunhut for providing his routine to
automatically adjust masks. This research has made use of the SIMBAD database
operated at CDS, Strasbourg (France), and of NASA's Astrophysics Data System
(ADS). 

\bibliographystyle{mn2e}
\bibliography{articles}

\bsp

\label{lastpage}

\end{document}